\documentclass[reprint,superscriptaddress,amsmath,amssymb,aps,prb]{revtex4-1}

\usepackage{array}
\usepackage{multirow}
\usepackage{dcolumn}
\usepackage{hhline}
\usepackage[normalem]{ulem}
\newcolumntype{P}[1]{>{\centering\arraybackslash}m{#1}}

\usepackage{graphicx}
\usepackage{bm}
\usepackage{natbib}
\usepackage{color}

\newcommand{\ket}{\rangle}
\newcommand{\bra}{\langle}

\newcommand{\+}{\dagger}

\newcommand{\dd}{\mathrm{d}}

\newcommand{\nablab}{\boldsymbol{\nabla}}

\newcommand{\pinf}{\scriptscriptstyle{+\infty}}

\newcommand{\tr}{\mathrm{Tr}}
\newcommand{\im}{\mathfrak{Im}}

\newcommand{\dbar}{\overset{\rotatebox{90}{${\!\scriptscriptstyle\parallel}$}}}

\newcommand{\pos}{\mathbf{r}}
\newcommand{\ef}{\mathbf{E}}

\newcommand{\dip}{\boldsymbol{\mu}}
\newcommand{\rot}{\nablab\times}

\newcommand{\green}{\dbar{\mathbf{G}}}

\newcommand{\vecf}{\mathbf{f}}

\newcommand{\intzp}{\int_0^{\pinf}\hspace{-1em}}

\newcommand{\op}{\widehat}
\newcommand{\vac}{\mathbf{0}}

\newcommand{\hsig}{\op{\sigma}}

\begin{document}
\preprint{APS/123-QED}

\title{Comparative study of plasmonic antennas for strong coupling and quantum nonlinearities with single emitters}

\author{Benjamin Rousseaux}
\email{benjamin.rousseaux@chalmers.se}
\affiliation{Department of Physics, Chalmers University of Technology, 412 96 G\"oteborg, Sweden}
\affiliation{Department of Microtechnology and Nanoscience - MC2, Chalmers University of Technology, 412 96 G\"oteborg, Sweden}
\author{Denis G. Baranov}
\affiliation{Department of Physics, Chalmers University of Technology, 412 96 G\"oteborg, Sweden}
\author{Mikael K\"all}
\affiliation{Department of Physics, Chalmers University of Technology, 412 96 G\"oteborg, Sweden}
\author{Timur Shegai}
\affiliation{Department of Physics, Chalmers University of Technology, 412 96 G\"oteborg, Sweden}
\author{G\"oran Johansson}
\affiliation{Department of Microtechnology and Nanoscience - MC2, Chalmers University of Technology, 412 96 G\"oteborg, Sweden}

\begin{abstract}
Realizing strong coupling between a single quantum emitter (QE) and an optical cavity is of crucial importance in the context of various quantum optical applications. While Rabi splitting of single quantum emitters coupled to high-$Q$ diffraction limited cavities have been reported in numerous configurations, attaining single emitter Rabi splitting with a plasmonic nanostructure is still elusive.
Here, we establish the analytical condition for strong coupling between a single QE and a plasmonic nanocavity and apply it to study various plasmonic arrangements that were shown to enable Rabi splitting. We investigate numerically the optical response and the resulting Rabi splitting in metallic nanostructures such as bow-tie nanoantennas, nanosphere dimers and nanospheres on a surface and find the optimal geometries for emergence of the strong coupling regime with single QEs. We also provide a master equation approach to show the saturation of a single QE in the gap of a silver bow-tie nanoantenna. Our results will be useful for implementation of realistic quantum plasmonic nanosystems involving single QEs.
\end{abstract}
\maketitle
\section{Introduction}

Interaction between electromagnetic modes of a cavity and a two-level quantum emitter (QE), in the most simple picture described by the Jaynes-Cummings (JC) Hamiltonian, is responsible for a rich variety of peculiar quantum optical effects~\cite{JaynCumm63,Scully,Dutra}. The Rabi frequency $\Omega$, which in this framework determines the strength of interaction between light and matter, is the crucial parameter determining the behavior of a strongly coupled system.
The weak light-matter interaction, realized when the Rabi frequency is small compared to rates of dissipative processes, manifests itself in the Markovian dynamics of the system characterized by the irreversible spontaneous decay of the QE~\cite{Weisskopf,Allen}. It can be further accelerated by increasing the local density of optical states (LDOS), e.g. using a cavity~\cite{Purcell1946,Pelton,MDPurcell}.

Strong coupling, on the other hand, is a special regime of light-matter interaction, emerging when the Rabi frequency exceeds the rates of incoherent processes~\cite{KimbLynn03,Khitrova2006, Torma2015,baranov2017novel}. In this regime of light-matter interaction the photonic and matter components of the system can no longer be treated as separate entities, as they form the polaritonic states (sometimes referring to as the \emph{dressed states}) with their eigenenergies being separated by the Rabi splitting of $\Omega$~\cite{Khitrova2006,Fink2008, Bishop2009, Kasprzak2010}. Such evolution of the system spectrum modifies its response and dramatically affects transport~\cite{Transport,Conductance} and chemical~\cite{Angew2012,Angew2016,Chemistry2016,Galego2016} properties.
Strong light-matter coupling is particularly interesting in the single emitter limit, when unique features of the Jaynes-Cummings ladder enable ultrafast and single-photon optical nonlinearities~\cite{Birnbaum2005,Englund2007}. 

While strong coupling between high-$Q$ dielectric cavities and single emitters such as atoms, quantum dots, and superconducting qubits has been the subject of intense research~\cite{Reithmaier2004,Yoshie2004,Kimble04,Peter2005}, much less progress has been made in realization of strong coupling between single emitters and \emph{plasmonic} nanostructures. The ability to strongly couple a single QE to a nanoscale plasmonic antenna would be extra beneficial for quantum information processing applications \cite{Tame,Dzsotjan,Volz2012}. Although observations of Rabi splitting with ensembles of quantum emitters coupled to plasmonic structures have been widely reported~\cite{Fofang2011,Schlather2013,Rodriguez2013,Cacciola2014, Zengin2015,Torma2015,ZengSheg13}, only a few recent reports claimed observation of Rabi splitting with a single QE~\cite{Shvets,Santhosh2016,Chikkaraddy2016,Hua2017,GrosHech18}. Achieving prominent and robust splittings in plasmonic structures is hindered mostly by low $Q$-factors of such nanocavities -- a problem that has been suggested to deal with via structuring environment of the emitter \cite{YunFengEnhancing,gurlek2017manipulation}.

Reaching the regime of strong coupling with a single QE is challenging. Indeed, several previous theoretical studies have described a related but less demanding regime of interaction, i.e., the emergence of Fano resonance in single QE-plasmon systems \cite{WuPelt10,ShahGray13,ChenAgio13}. Such phenomena can also lead to single photon nonlinearities at the nanoscale with the saturation of the QE when the system is driven strongly enough. However, Fano interference requires both high quantum yield of the nanoantenna and low internal loss of the QE, thus low temperature setups. Strongly coupled single QE-plasmon systems would enable the observation of single photon nonlinearities at room temperature, as the QE drastically affects the optical response of the plasmonic structure. Studies reporting strong coupling between a single QE and plasmonic cavities have revealed characteristic values for both the Purcell factors and the QE dipole moments involved in this regime \cite{TrugHohe08,SavaBorg10,ManjNord11,HartLi15}.

In this paper, we systematically address the problem of strong coupling with a single QE, within a quantum description of plasmonic resonances.
We establish the analytical condition for strong coupling of between a single QE and a plasmonic nanocavity and analyze various plasmonic nanostructures enabling Rabi splitting. We study numerically optical response and the resulting Rabi splitting in metallic nanostructures such as bow-tie nanoantennas, nanosphere dimers and nanospheres on a surface and find the optimal geometries for emergence of the strong coupling regime with single QEs corresponding to several existing materials. We also discuss the impact of multiple modes of a plasmonic resonator on the strong coupling regime, and the trade-off between radiative and non-radiative processes.

\begin{figure}[b]
\includegraphics[width=.9\columnwidth]{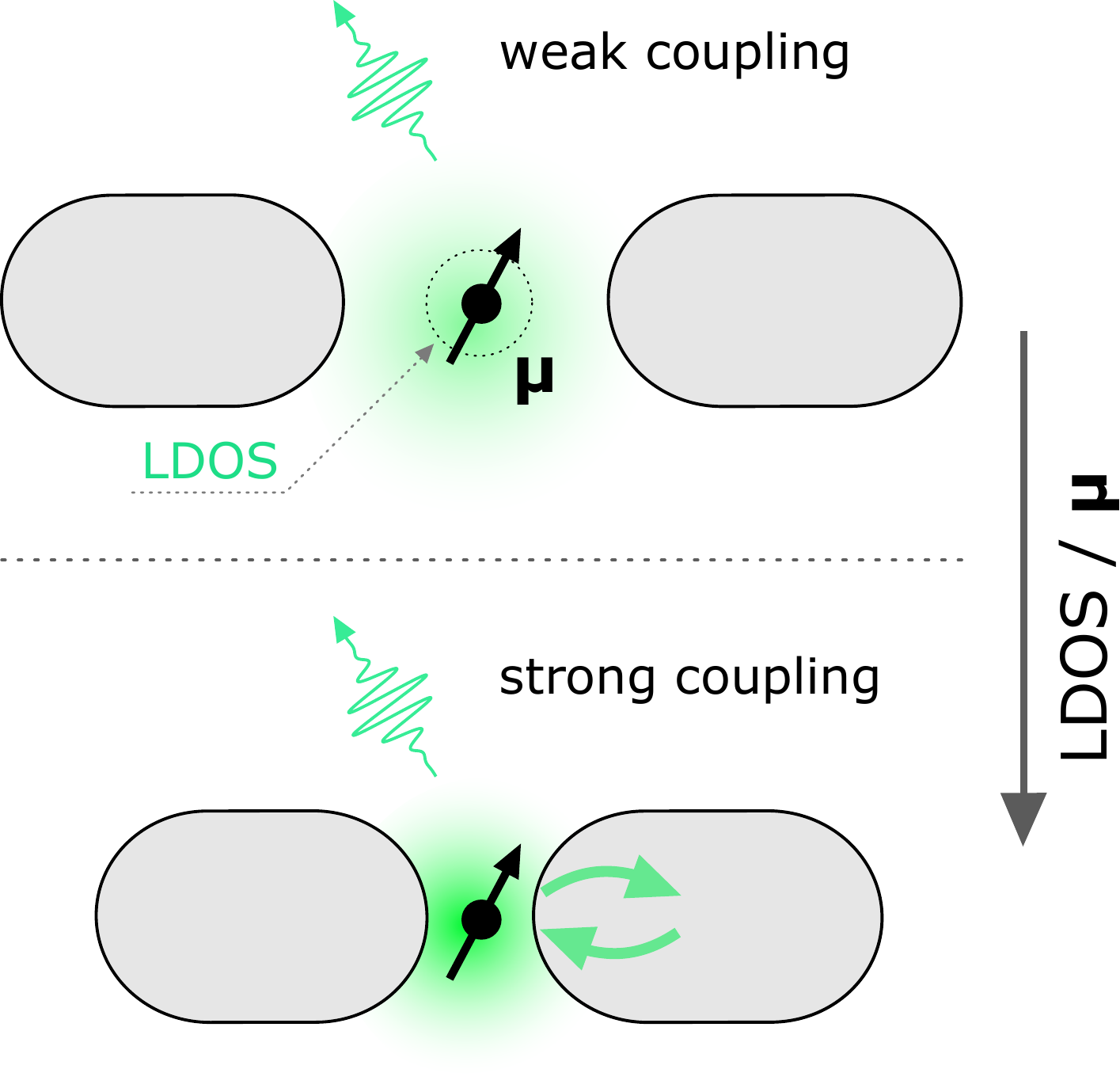}
\caption{Schematic of the structure under study: a two-level system modeling a quantum emitter positioned in hotspot of a nanoantenna. Upon increase of the LDOS or the QE transition dipole moment, the system exhibits transition from the weak to strong coupling regime, when Rabi oscillations take over spontaneous emission.}
\label{fig0}
\end{figure}

\section{Quantum description and strong coupling regime}
\subsection{Local density of states (LDOS) and Purcell factor}
In this section, we develop the theoretical framework for the description of a single QE coupled to a nanoscale structure. This framework is based on the description of spontaneous emission in a structured environment, which is originally pictured by the Purcell effect, as the QE decays faster into an environment with a higher number of surrounding modes. We derive the quantization procedure leading to a set of specific cavity modes associated with localized surface plasmon resonances, enabling the construction of the master equation for the dynamics, and derive the threshold condition for local density of states (LDOS) and the QE dipole moment, upon which strong coupling occur between the QE and the cavity modes (see Fig. \ref{fig0}).

We start with the formulation of the LDOS, which is a space-dependent quantity describing how fast quantum emitters can decay in localized field modes. It is determined with the Green's tensor of the electric field contribution, corresponding to the solution of the Maxwell-Helmholtz equation:
\begin{align}
\Big(\rot\rot - \frac{\omega^2}{c^2}\epsilon(\pos,\omega)\Big)\green_\omega(\pos,\pos') = \dbar{\rm I}\delta(\pos - \pos'),
\end{align}
where $\epsilon(\pos,\omega)$ is the dielectric function and $\dbar{\rm I}$ is the $3\times 3$ identity matrix. For an emitter whose transition dipole moment is along the unit vector $\mathbf{n}$, the LDOS is:
\begin{align}
\label{LDOS}
\rho_\mathbf{n}(\pos_E,\omega) = \frac{6\omega}{\pi c^2}\mathbf{n}\cdot\im\left\{\green_\omega(\pos_E,\pos_E)\right\}\mathbf{n},
\end{align}
where $\pos_E$ is the position of the emitter and $\omega$ is the angular frequency. We also consider the quantity defined through the relation:
\begin{align}
F_P(\omega) = \frac{\rho_\mathbf{n}(\pos_E,\omega)}{\rho_\mathbf{n}^0(\pos_E,\omega)},
\label{Purcellf}
\end{align}
which is the Purcell factor at frequency $\omega$ (and $\rho_\mathbf{n}^0(\pos_E,\omega)$ is the LDOS in the absence of the nanoantenna). We note that the Purcell factor is a property of the nanoantenna alone, hence it is independent of the QE dipole moment. It is seen from \eqref{LDOS} that the knowledge of the Green's tensor $\green_\omega(\pos_E,\pos_E)$ provides the LDOS and the Purcell factor. Green's dyadic can be obtained in two manners: using analytical derivations, and using numerics. In the case of spherical and spheroidal geometries (e.g. for spherical or prolate spheroidal nanoparticles), the Green's tensor can be obtained analytically \cite{LiYeo94,LiLeon02}. Nevertheless, for more complicated geometries, one has to rely on numerical calculations \cite{Sauvan2013}. The tools for getting the Green's function of arbitrary geometries are provided by finite-difference time domain (FDTD) softwares.
        
  \subsection{Quantum description of a single emitter coupled to plasmonic modes}

    \subsubsection{General Green's tensor approach}

To understand the different coupling regimes, we use a quantum description and derive an effective Hamiltonian \cite{RousGuer16,DzsoGuer16,VargCola16} whose structure is completely analogous to cavity QED Hamiltonians. The derivation is based on a first-principle method corresponding to the coupling of a single two-level system with a reservoir of harmonic oscillators: the Fano diagonalization \cite{KnolWels01,Phil10}. In this description, the electric field is toggled by creation and annihilation operators $\op{\vecf}_\omega^\+(\pos),\op{\vecf}_\omega(\pos)$ obeying the commutation relations:
\begin{align}
\big[\,\op{\vecf}_\omega(\pos),\op{\vecf}_{\omega'}^\+(\pos')\big] = \delta(\pos - \pos')\delta(\omega - \omega').
\end{align}
In other words, the environment of the quantum emitter corresponding to the nanoparticles and the surrounding free space is diagonalized, and described by these global operators. In addition, the geometry of the structure considered is reflected in the use of the Green's tensor $\green_\omega(\pos,\pos')$ in the expression of the electric field:
\begin{subequations}
\label{ef}
\begin{align}
\op{\ef}(\pos) &= \intzp\dd\omega\Big(\op{\ef}^{(+)}_\omega(\pos) + \op{\ef}^{(-)}_\omega(\pos)\Big),\\
\op{\ef}^{(+)}_\omega(\pos) &= i\sqrt{\frac{\hbar}{\pi\epsilon_0}}\frac{\omega^2}{c^2}\int\dd^3r'\sqrt{\epsilon''_\omega(\pos')}\green_\omega(\pos,\pos')\cdot\op{\vecf}_\omega(\pos'),\\
\op{\ef}^{(-)}_\omega(\pos) &= [\op{\ef}^{(+)}_\omega(\pos)]^\+,
\end{align}
\end{subequations}
where we introduced the imaginary part of the dielectric function, $\epsilon''_\omega(\pos) = \im\{\epsilon(\pos,\omega)\}$, which contains the information about the optical properties of the geometry. The calculation of the Green's tensor along with the boundary conditions projects the general structure of the field into a subspace corresponding to the specific choice of the geometry. In this consideration the basis of the creation-annihilation operators $\op{\vecf}_\omega^\+(\pos),\op{\vecf}_\omega(\pos)$ is also projected onto a subspace of electromagnetic modes, and the field Hamiltonian reduces to the basis:
\begin{multline}
\op{H}_\text{field} = \intzp\dd\omega\,\hbar\omega\int\dd^3r\,\op{\vecf}_\omega^\+(\pos)\op{\vecf}_\omega(\pos)\\
\longrightarrow \sum_{\boldsymbol{\eta}}\intzp\dd\omega\,\hbar\omega\,\op{a}_{\boldsymbol{\eta}}^\+(\omega)\op{a}_{\boldsymbol{\eta}}(\omega),
\end{multline}
where here the creation-annihilation operators $\op{a}_{\boldsymbol\eta}^\+(\omega),\op{a}_{\boldsymbol\eta}(\omega)$ toggle excitations of localized surface plasmon cavity modes labeled by the general index $\boldsymbol\eta$ at frequency $\omega$. For the new operators to obey the commutation relation $[\op{a}_{\boldsymbol\eta}(\omega),\op{a}_{\boldsymbol\eta'}^\+(\omega')] = \delta_{\boldsymbol\eta\boldsymbol\eta'}\delta(\omega - \omega')$, it is required that all cavity modes $\boldsymbol\eta$ are orthogonal to each other. As an example, in the case of an emitter close to a nanosphere, this general index reduces to a harmonic index $n = 1,2,...$ whose values are associated with specific mode geometries: $n=1$ is the dipolar mode, $n=2$ the quadrupolar mode, etc \cite{RousGuer16,DzsoGuer16}. The coupling of the plasmonic field and the emitter is introduced with a dipole coupling term $-\op{\dip}\cdot\op{\ef}(\pos_E)$, where the QE is a point-like two-level quantum system with a finite dipole moment $\mu$, and the total rotating wave approximation (RWA) Hamiltonian reads:
\begin{multline}
\label{Hamiltonian}
\op{H} = \hbar\omega_0\hsig_+\hsig_- + \sum_{\boldsymbol{\eta}}\intzp\dd\omega\,\hbar\omega\,\op{a}_{\boldsymbol{\eta}}^\+(\omega)\op{a}_{\boldsymbol{\eta}}(\omega)\\
+ i\hbar\sum_{\boldsymbol\eta}\intzp\dd\omega\Big(\kappa_{\boldsymbol\eta}(\omega)\op{a}_{\boldsymbol\eta}^\+(\omega)\hsig_- - \text{h.c.}\Big).
\end{multline}
In the above expression, $\omega_0$ denotes the transition frequency of the two-level emitter, $\hsig_- = |g\ket\bra e|, \hsig_+ = \hsig_-^\+$ are the lowering and raising operators of the transition, and $\kappa_{\boldsymbol\eta}(\omega)$ is the emitter-mode coupling, associated with a specific mode $\boldsymbol\eta$. The latter contains the Green's tensor through the expression:
\begin{align}
\label{coupl}
|\kappa_{\boldsymbol\eta}(\omega)|^2 = \frac{1}{\hbar\pi\epsilon_0}\frac{\omega^2}{c^2}\dip\cdot\im\left\{\green_{\boldsymbol\eta,\omega}(\pos_E,\pos_E)\right\}\dip,
\end{align}
with $\green_{\boldsymbol\eta,\omega}(\pos_E,\pos_E)$ being the Green's tensor corresponding to mode $\boldsymbol\eta$. We see that the square modulus of the emitter-field coupling is proportional to the LDOS \eqref{LDOS}, for a given mode. The knowledge of the latter is then essential for the understanding of the quantum description.

\subsubsection{Single mode effective model}

In cavity QED, the description of the cavity field is often, to a good approximation, taken to be a single mode. In quantum plasmonics, however, this is often not the case and one has to include many plasmonic modes in the model, as is the case for spherical nanoparticles. The single mode approach can nevertheless be a useful tool for understanding the different coupling regimes.

Let us consider a single mode $\boldsymbol\eta$ interacting with a QE.
The corresponding emitter-mode coupling $\kappa(\omega)$ has a Lorentzian shape \cite{RousGuer16,DzsoGuer16}:
\begin{align}
\label{lor_coupl}
\kappa(\omega)\approx \sqrt{\frac{\gamma_{\rm cav}}{2\pi}}\frac{g}{\omega - \omega_c + i\frac{\gamma_{\rm cav}}{2}},
\end{align}
where $\omega_c$ is the cavity mode frequency, $\gamma_{\rm cav}$ is the cavity decay rate being equal to the full width of its scattering spectrum, and $g$ is the usual Jaynes-Cummings interaction constant:
\begin{align}
\label{g}
g = \sqrt{\frac{\pi\omega_r\gamma_{\rm cav}\rho_\mathbf{n}(\pos_E,\omega_r)}{12\hbar\epsilon_0}}\mu.
\end{align}
Solving the time-dependent Schr\"odinger equation with Hamiltonian \eqref{Hamiltonian} and using the Lorentzian profile \eqref{lor_coupl}, it can be shown that the full Hamiltonian reduces to the effective Hamiltonian:
\begin{align}
\label{H_eff}
\op{H}_\text{eff} = \left(\begin{array}{cc}\omega_0&g\\g&\omega_c-i\frac{\gamma_{\rm cav}}{2}\end{array}\right),
\end{align} 
which is expressed in the single excitation basis $\{|e,0\ket,|g,1\ket\}$ where only one photon is exchanged via the $|e\ket \leftrightarrow|g\ket$ transition. This effective Hamiltonian is non-Hermitian due to the loss terms on the diagonal, as this arises when coupling a discrete state coupled to a continuum of modes. It is also equivalent to take only the Hermitian part of this Hamiltonian and write a Lindblad master equation for the time evolution of the state \cite{Garraway96,DaltGarra01}:
\begin{multline}
\label{meqn}
\dot{\op{\varrho}} = -i\big[\op{H}_\text{JC},\op{\varrho}(t)\big]\\ + \gamma_\text{cav}\Big(\op{a}\op{\varrho}(t)\op{a}^\+ - {\textstyle\frac{1}{2}}\op{a}^\+\op{a}\op{\varrho}(t)- {\textstyle\frac{1}{2}}\op{\varrho}(t)\op{a}^\+\op{a}\Big),
\end{multline}
where $\op{\varrho}(t)$ is the density operator for the state of the QE-plasmon system, $\op{a}$ is the annihilation operator of the plasmon mode and $\op{H}_\text{JC} = \omega_0\hsig_+\hsig_- + \omega_c\op{a}^\+\op{a} + g(\op{a}^\+\hsig_- + \op{a}\hsig_+)$ is the Jaynes-Cummings Hamiltonian.


\section{Strong coupling condition and comparative study}
\subsection{Analytical condition}
\label{analy_cond}
The dynamics of a single QE coupled to a plasmonic mode is parametrized by three quantities: the JC interaction constant $g$, the QE free space decay rate $\gamma_0$ and the cavity mode decay rate $\gamma_{\rm cav}$.
In the weak coupling regime, when $g \ll \gamma_{\rm cav}$, the QE experiences spontaneous emission: the probability to find the QE in the excited state $P_e (t)$ decays exponentially in time with the rate $\gamma_0 F_P(\omega_0)$,
where $F_P$ is the well-known Purcell factor, which reflects acceleration of the spontaneous emission due to enhanced local density of states~\cite{Purcell1946,Pelton,MDPurcell}.
In the single-mode approximation and assuming that the QE is resonant with the cavity mode, $\omega_c = \omega_0$, the Purcell factor can be expressed through the interaction constant via \cite{Scully}
\begin{equation}
{F_P} = \frac{4g^2}{\gamma _{\rm cav}\gamma _0}.
\end{equation}

As the interaction constant increases, $g \sim \gamma_{\rm cav}$ or even $g > \gamma_{\rm cav}$, single photons start to oscillate between the emitter and the mode, giving rise to non-Markovian dynamics of the emitter population $P_e (t)$, which is a signature of the strong coupling regime.
Mathematically, strong coupling occurs whenever separation between the two eigenvalues of Hamiltonian \eqref{H_eff} exceeds $(\gamma_{\rm cav}+\gamma_0)/2$. In general, they should even exceed $(\gamma_{\rm cav} + \gamma)/2$, where $\gamma = \gamma_0 + \gamma_{\rm inh}$ is the total decay rate of the transition, accounting for both free space and inhomogeneous decay. For simplicity, we will consider $\gamma_{\rm inh} = 0$ in this work. 
Comparing the two eigenvalues of \eqref{H_eff} (assuming that the QE is resonant with the nanoantenna mode, $\omega_0 = \omega_r$, and $\gamma_{\rm cav}\gg\gamma_0$), we find that the onset of strong coupling corresponds to the interaction constant $g = \gamma_{\rm cav}/(2\sqrt{2})$. Conversely, estimating the Purcell factor $F_P$ at the onset of strong coupling, we obtain a convenient threshold condition for strong coupling in terms of the Purcell factor:
\begin{align}
F_P(\omega_0) \geqslant \frac{\gamma_{\rm cav}}{2\gamma_0}.
\label{SC}
\end{align}
Typical values of the plasmon lifetime and the QE lifetime lie on the scale of 10 fs and 1 ns, respectively. The latter equation thus suggests that $F_P$ should be at least of the order of $10^5$ in order to reach strong coupling with a single QE coupled to a realistic plasmon resonator.

In order to reach strong coupling, the interaction constant $g$ should be increased. This can be done in two different manners: either by increasing the transition dipole moment $\mu$ of the QE or by increasing the Purcell factor $F_P$.
The set of available QEs is often limited to a few options. Much greater flexibility is, however, offered by  designing the optical cavity and maximizing the Purcell factor.

Expression (\ref{SC}) allows one to estimate the threshold magnitude of the dipole moment $\mu_\text{s.c.}$ of a single QE which is required to reach strong coupling with a given nanoantenna with some value of $F_P$.
To do so, we note that the free space spontaneous emission rate $\gamma_0$ in Eq. (\ref{SC}) depends on $\mu$; expressing $\mu$ from this formula yields us the threshold dipole moment magnitude:
\begin{align}
\mu_\text{s.c.} = \sqrt{\frac{3\hbar\pi\epsilon_0c^3\gamma_{\rm cav}}{2\omega_0^3F_P}}.
\label{musc}
\end{align}
This expression can be used to find the optimal geometry for reaching the strong coupling regime with a single emitter: one needs to calculate the Purcell factor and the cavity decay rate and check if the resulting values are enough to guarantee strong coupling with a given emitter.

\begin{figure}
\begin{center}
\includegraphics[width=.9\columnwidth]{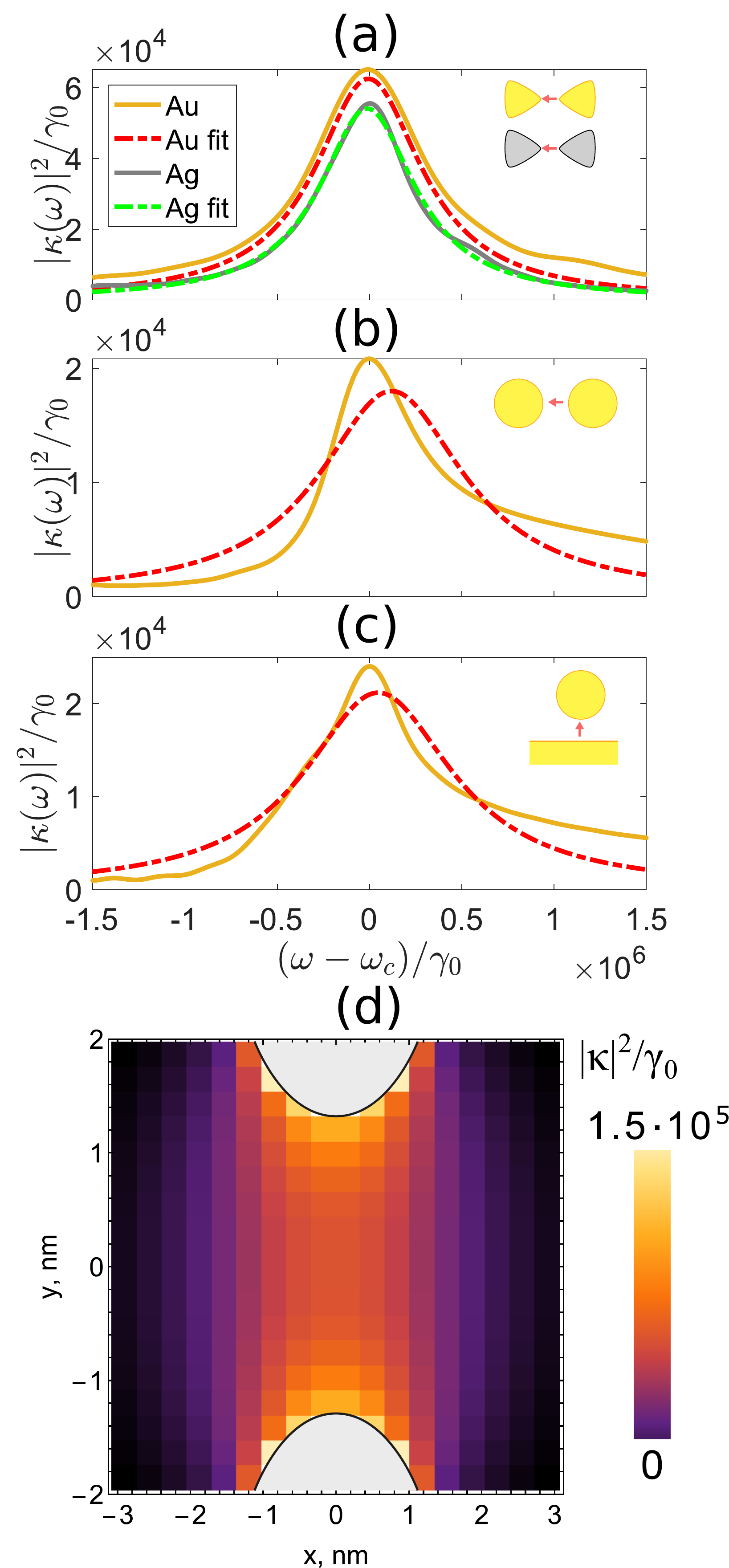}
\caption{(a)-(c) FDTD calculation of the QE-plasmon coupling $|\kappa(\omega)|^2$ proportional to the LDOS at the center of the gap (solid blue) and Lorentzian fit of the considered modes (dash-dotted red). The calculation is done for four different geometries. (a) Au and Ag  bow-tie nanoantennas. (b) Au nanosphere dimer. (c) Au NPoS. The permittivity of Ag and Au was adopted from Ref. \citenum{johnson1972optical}. For bow-tie nanoantennas, triangles with 90 nm side length, 50 nm base, and 10 nm thickness were used. For NPoS, nanospheres with 50 nm diameter were used. For dimer, 30 nm nanospheres were used. The gap in all cases was set to 3 nm.
(d) Spatial map of the total Purcell factor for an Ag bow tie nanoantenna with a 3 nm gap at resonance ($\sim 750$ nm) for an electric dipole oriented along the gap.}
\label{lorentzian_fit}
\end{center}
\end{figure}

\subsection{Comparative study of the nanoantenna geometry}
Now we employ the condition \eqref{musc} to study the feasibility of reaching the strong coupling regime in various plasmonic nanoantennas with single QEs. We will consider three nanoantenna types: (i) dimer of two nanospheres, (ii) nanoparticle on a surface (NPoS), and (iii) bow-tie nanoantenna. We also consider two materials from which antennas are made: gold (Au) and silver (Ag) for the bow-tie (resonances in the other two Ag antennas are strongly shifted to the UV region which makes them less interesting). Nanometer gaps in these antennas ensure deeply subwavelength mode volumes and high Purcell factors lowering the required transition dipole moment. To study the coupling in these nanoantennas, we calculated numerically the LDOS spectra using the Lumerical FDTD solver. The maximum of each spectrum is then identified as the plasmonic mode, and a corresponding QE resonant with that mode is assumed in further calculations.

The LDOS spectra for each geometry considered here are presented in Fig. \ref{lorentzian_fit} for a 3 nm gap in all cases. It is seen that the LDOS of bow-tie nanoantennas around the bright mode is a single Lorentzian, but this is not the case for NPoS and sphere dimers, due to the presence of other strongly non-radiative modes. The calculation shown in fig. \ref{lorentzian_fit} is then done for different gap sizes, using the same geometries.

\begin{figure}
\begin{center}
\includegraphics[width=1\columnwidth]{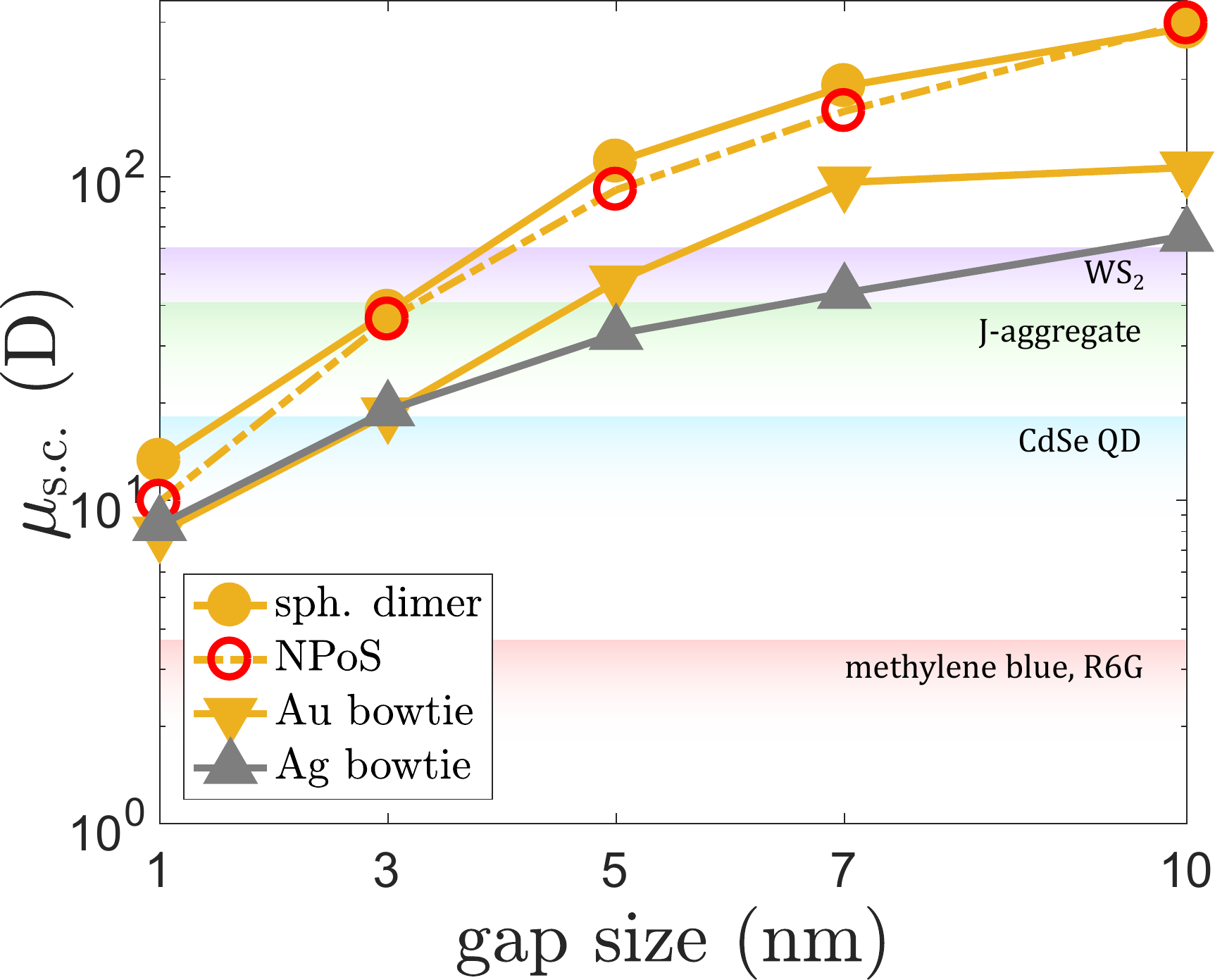}
\caption{\small Minimal dipole moment \eqref{musc} for reaching the strong coupling regime, in debyes (D). The calculations are performed for Au and Ag bow-ties, Au NPoS and Au nanosphere dimer.}
\label{min_dip}
\end{center}
\end{figure}

The calculated spectra are then fitted with an effective Lorentzian function \eqref{lor_coupl} parametrized by the width $\gamma_{\rm cav}$, the Jaynes-Cummings coupling $g$ and the resonance frequency $\omega_c$.
Replacing the full LDOS by a single Lorentzian fit works well for the bow-tie nanoantennas, Fig. \ref{lorentzian_fit}(a,b).
For the case of NPoS and nanosphere dimers, however, the total LDOS does not feature a Lorentzian behaviour. However, for simplicity we fit the LDOS with a Lorentzian peak using the least square method, as we are interested in the global coupling strength between the nanoantenna and the QE.

The LDOS and hence the coupling strength vary with the position at which the QE is located with respect to the nanoantenna. This is illustrated in Fig. \ref{lorentzian_fit}(e) for the Ag bow-tie nanoantenna with 3 nm gap for a QE's dipole moment oriented along the gap. The Purcell factor reaches the maximal values close to the antenna edges, and quickly decays away from the edges. This plot also emphasizes that positioning of a QE exactly in the center of a gap is neither necessary nor optimal for ensuring the highest LDOS and interaction constant. 
Having performed the fitting procedure for each LDOS spectrum, we now calculate the threshold dipole moment $\mu_\text{s.c.}$ of a resonant QE to reach the strong coupling regime for each nanoantenna.
The results are presented in Fig. \ref{min_dip}.
Expectedly, the threshold dipole moment quickly decreases with shrinking gap for all antennas. Among all studied geometries, the bow-tie antennas exhibit the lowest threshold dipole moment, which is due to the lowest resonance width $\gamma_{\rm cav}$ and a tightly localized electric field at the tips of the bow-tie nanoantenna. On the same plot we show horizontal bars corresponding to a few QEs often employed in the studies of strong light-matter interaction. The key observation is that QEs such as excitons in monolayer WS{$_2$} and J-aggregates may allow us to reach strong coupling in the single emitter limit with bow-tie nanoantennas with gaps as large as 5 nm. As the gap shrinks down to 1 nm, all four systems exhibit comparable response, with simple organic molecules such as methylene blue being on the edge of strong coupling.

Note that for gaps smaller than 5 nm non-local response of metal may become pronounced, whereas our calculations of the coupling strength are based on the local model. This may have an unfavorable effect on the coupling strength, as recent theoretical efforts suggest \cite{jurga2017plasmonic}.

\begin{table}
  \begin{tabular}{| P{2.cm} | P{1.1cm} |P{1.7cm} | P{1.1cm} | P{1.1cm} |}
\hline
Nanoantenna  & NPoS &  Nanosphere dimer & Au bow-tie & Ag bow-tie  \\ \hline
$\eta$   &  0.0015 & 0.0026 & 0.2796 & 0.4513  \\ \hline
\end{tabular}
\caption{Quantum yield $\eta$ of an emitter placed in the center of the gap tuned to the antenna resonance for the four studied nanoantennas.}
\label{QE}
\end{table}

\begin{figure}
\includegraphics[width=.7\columnwidth]{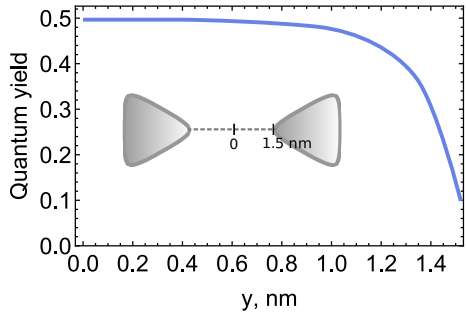}
\caption{Spatial dependence of the quantum yield of a resonant emitter interacting with a 3 nm gap Ag bow-tie nanoantenna. The emitter wavelength is 610 nm, the dipole moment is oriented in $y$ direction. the origin is chosen to be 0 and $y = 1.5$ nm corresponds to the tip of one nanoprism.}
\label{qu_yield}
\end{figure}

We further inspect the radiative properties of the four considered geometries. For the Rabi splitting to be clearly observable in the emission spectrum, the decay of a QE must be mostly radiative. This can be quantified by the quantum yield of the QE:
\begin{align}
\eta = \frac{\Gamma_\text{rad}}{\Gamma_\text{tot}}
\end{align}
where $\Gamma_\text{rad}$ is the radiative decay rate of the QE in the vicinity of the nanoantenna. The quantum yield of an emitter located at the gap center at the resonance wavelength for each nanoantenana is presented in Table \ref{QE}. The bow-tie geometry exhibits the best radiative properties among the studied antennas, orders of magnitude higher than NPoS and nanosphere dimer do. The spatial dependence of the quantum yield for the Ag bow-tie shown in Fig. \ref{qu_yield} suggests that the quantum yield is sensitive to the emitter position, but even very close to the metal edges the quantum yield reduces only down to 0.1, which is still 500 to 1000 times higher than the NPoS and nanosphere dimer value. Together with the result shown in Fig. \ref{lorentzian_fit}(e) it highlights the fact the precise positioning of a QE at a very specific site around the nanoantenna is not that crucial for reaching strong coupling as one might anticipate.

  \subsection{Single QE nonlinearity in a bow-tie nanoantenna}
\begin{figure}
\includegraphics[width=.8\columnwidth]{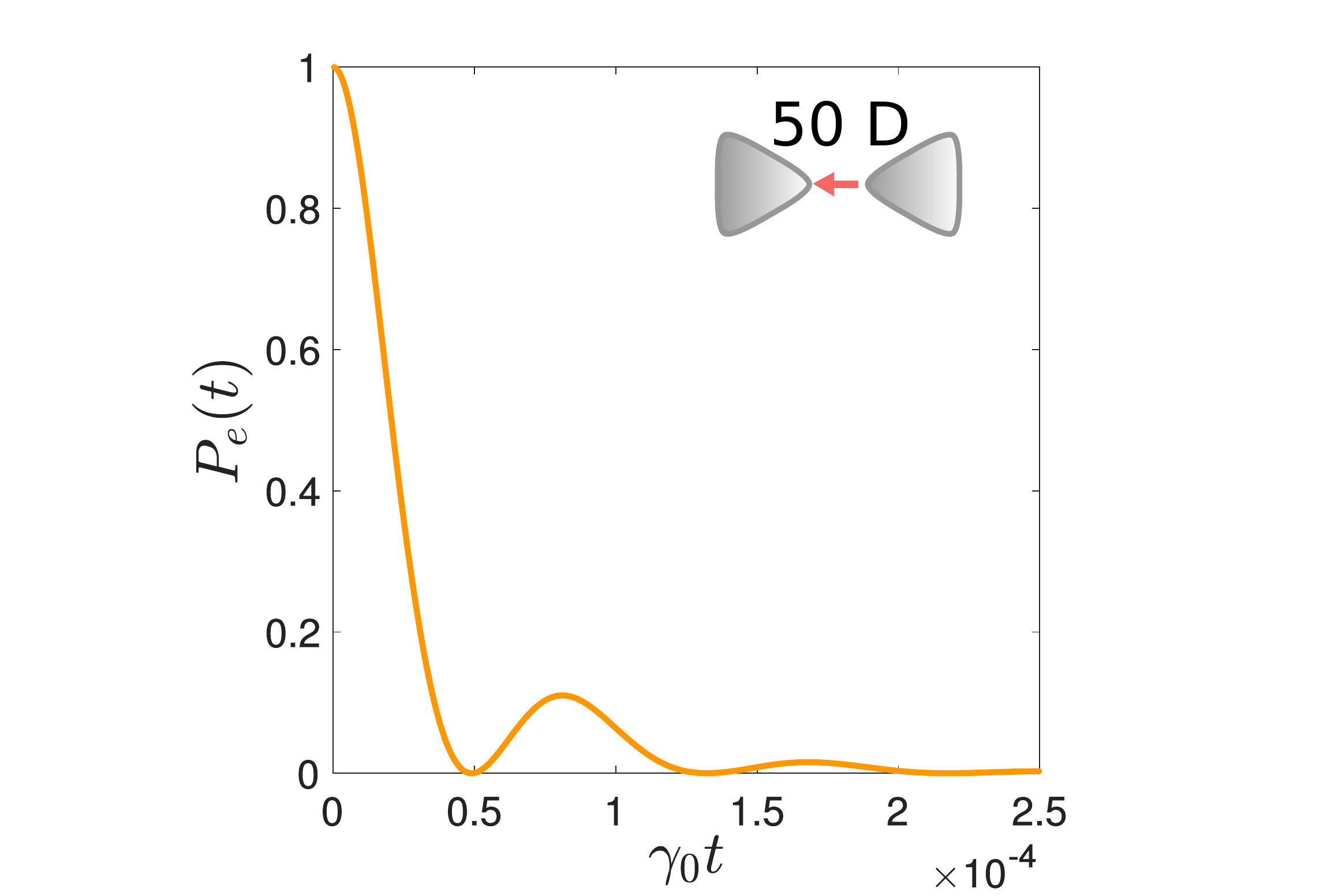}
\caption{Calculated temporal dynamics of population of a QE with 50 D transition dipole moment corresponding to the WS{$_2$} exciton. The  antenna geometry is tuned for its resonance to match the WS{$_2$} exciton wavelength of 610 nm.}
\label{fig5}
\end{figure}

Finally, to support our claims, we demonstrate the feasibility of strong coupling with a bow-tie nanoantenna and a single emitter via simulating the temporal dynamics of the emitter population. We choose a WS{$_2$} exciton with the transition dipole moment of 50 D \cite{SieGedi15,CuadSheg18} as a single QE and tune the Ag bow-tie resonance to the WS{$_2$} monolayer exciton transition wavelength of $\approx 610$ nm. Figure \ref{fig5}(a) shows the resulting emitter-mode coupling constant spectra of the Ag bow-tie with x nm side length, y nm height and 3 nm gap with the emitter located at the gap center. The corresponding temporal dynamics of the QE population presented in Fig. \ref{fig5}(b) clearly shows that at least one Rabi cycle occurs until the emitter irretrievably decays to the ground state.

While the calculation shown in Fig. \ref{fig5} is done considering a single excitation in the QE as an initial condition, that is $|\psi(0)\ket = |e\ket\otimes|\vac\ket$, the dynamics can also be determined using the steady-state solution of the master equation \eqref{meqn}. To do so we replace $\op{H}_{\rm JC}$ with a new Hamiltonian $\op{H}' = \op{H}_{\rm JC}' + \op{H}_{\rm pump}$ in a rotating frame oscillating with the pump frequency $\omega_P$, such that we have the effective RWA terms:
\begin{subequations}
\begin{align}
\op{H}_{\rm JC}' &= \Delta_0\hsig_+\hsig_- + \Delta_c\op{a}^\+\op{a} + g(\op{a}^\+\hsig_- + \op{a}\hsig_+),\\
\op{H}_{\rm pump} &= \frac{{\cal E}}{2}(\op{a} + \op{a}^\+),
\end{align}
\end{subequations}
where $\Delta_{0,c} = \omega_{0,c} - \omega_P$ and ${\cal E} = -\dip_{\rm n.a.}\cdot\ef_0/\hbar$ is the Rabi frequency associated with the pump laser field $\ef_P(t) = \ef_0\cos\omega_Pt$ and the effective transition dipole moment of the nanoantenna $\dip_{\rm n.a.}$. The steady state of the density operator is determined taking the time derivative in \eqref{meqn} being equal to zero and finding the non-trivial solutions of the equation:
\begin{align}
{\cal L}\varrho_{\rm s.s.} = 0,
\end{align}
${\cal L} = -i[\op{H}',\boldsymbol{\cdot}] + \gamma_\text{cav}(\op{a}\boldsymbol{\cdot}\op{a}^\+ - {\textstyle\frac{1}{2}}\op{a}^\+\op{a}\boldsymbol{\cdot}- {\textstyle\frac{1}{2}}\boldsymbol{\cdot}\op{a}^\+\op{a})$ being the superoperator acting on the density operator. The spectra are then computed using the average formula $\bra\op{A}\ket = \tr\{\varrho_{\rm s.s.}\op{A}\}$.

\begin{figure}
\includegraphics[width=\columnwidth]{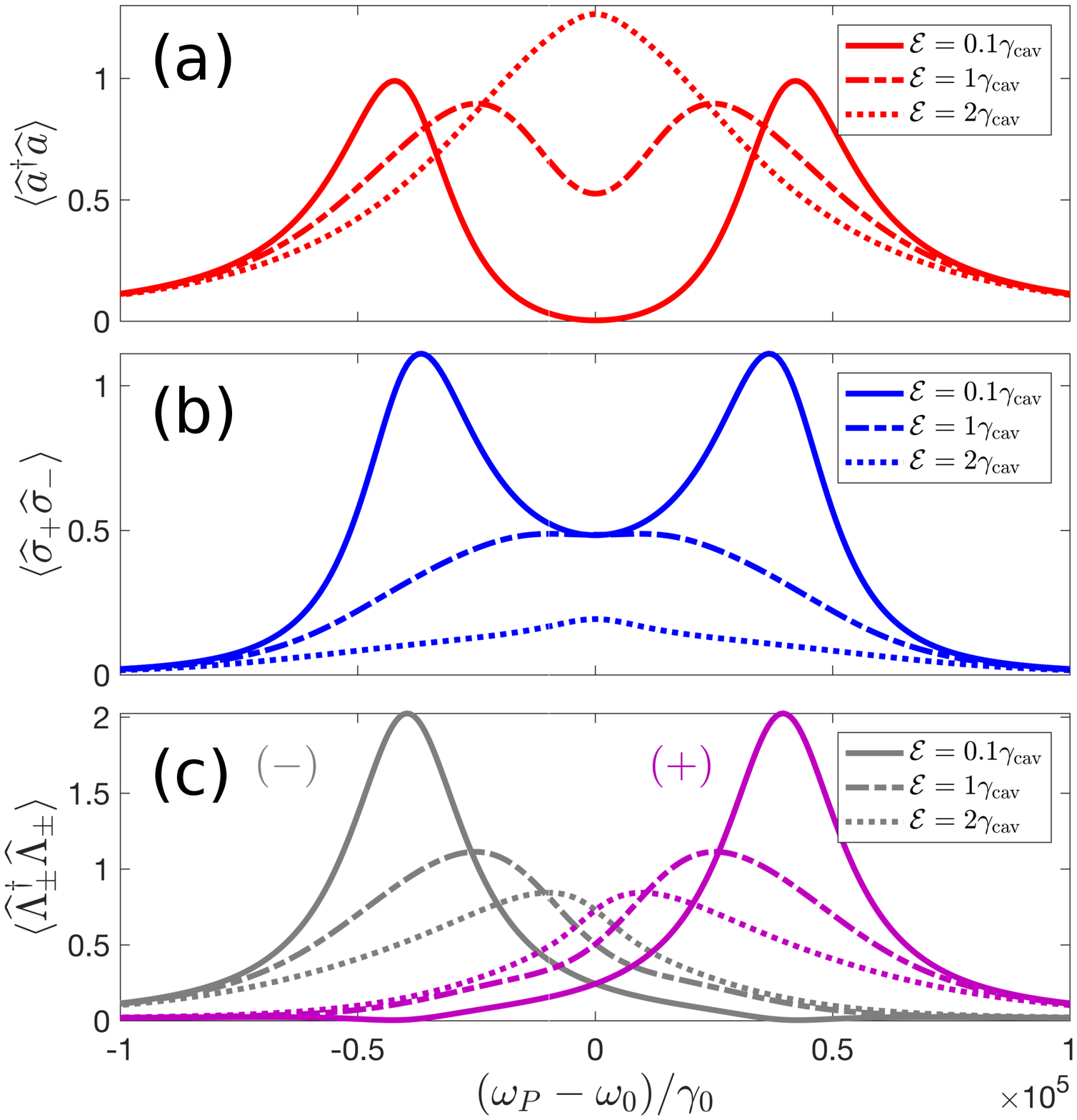}
\caption{Steady-state mode populations of a Ag bow-tie and single QE system precedingly studied in Fig. \ref{fig5}, versus pump frequency $\omega_P$, for different values of the pump Rabi frequency: ${\cal E} = 0.1\gamma_{\rm cav}$ (solid), $\gamma_{\rm cav}$ (dash-dotted) and $2\gamma_{\rm cav}$ (dotted). (a) Plasmonic resonance population $\bra\op{a}^\+\op{a}\ket$. (b) QE excitation population $\bra\hsig_+\hsig_-\ket$. (c) Polariton populations $\bra\op{\Lambda}_\pm^\+\op{\Lambda}_\pm\ket$ from the lower (grey) and upper (purple) polaritons. $y$-axes are in units of $1/{\cal E}^2$.}
\label{scmeqn}
\end{figure}

We evaluate the steady-state mode populations in Fig. \ref{scmeqn}, for three different quantities: the intensity of the cavity field $\bra\op{a}^\+\op{a}\ket$, the QE contribution $\bra\hsig_+\hsig_-\ket$, and their corresponding polaritonic mixture $\bra\op{\Lambda}_\pm^\+\op{\Lambda}_\pm\ket$, where the upper ($+$) and lower ($-$) polariton operators are given by:
\begin{align}
\op{\Lambda}_\pm = \frac{1}{\sqrt{2}}\big(\op{a}\pm\hsig_-\big).
\end{align}
The mode populations are calculated for three pump Rabi frequency values ${\cal E} = (0.1,1,2)\gamma_{\rm cav}$ and the saturation phenomenon is shown to happen for the two higher values. In particular, the plasmonic cavity Lorentzian lineshape is recovered for ${\cal E} = 2\gamma_{\rm cav}$, while the QE contribution is reduced and recombines as a single peak for the same value. We also see the same behavior in the polariton populations $\bra\op{\Lambda}_\pm^\+\op{\Lambda}_\pm\ket$, which we display as an interesting way to separate the lower and upper components in the dynamics.
\begin{figure}
\includegraphics[width=.9\columnwidth]{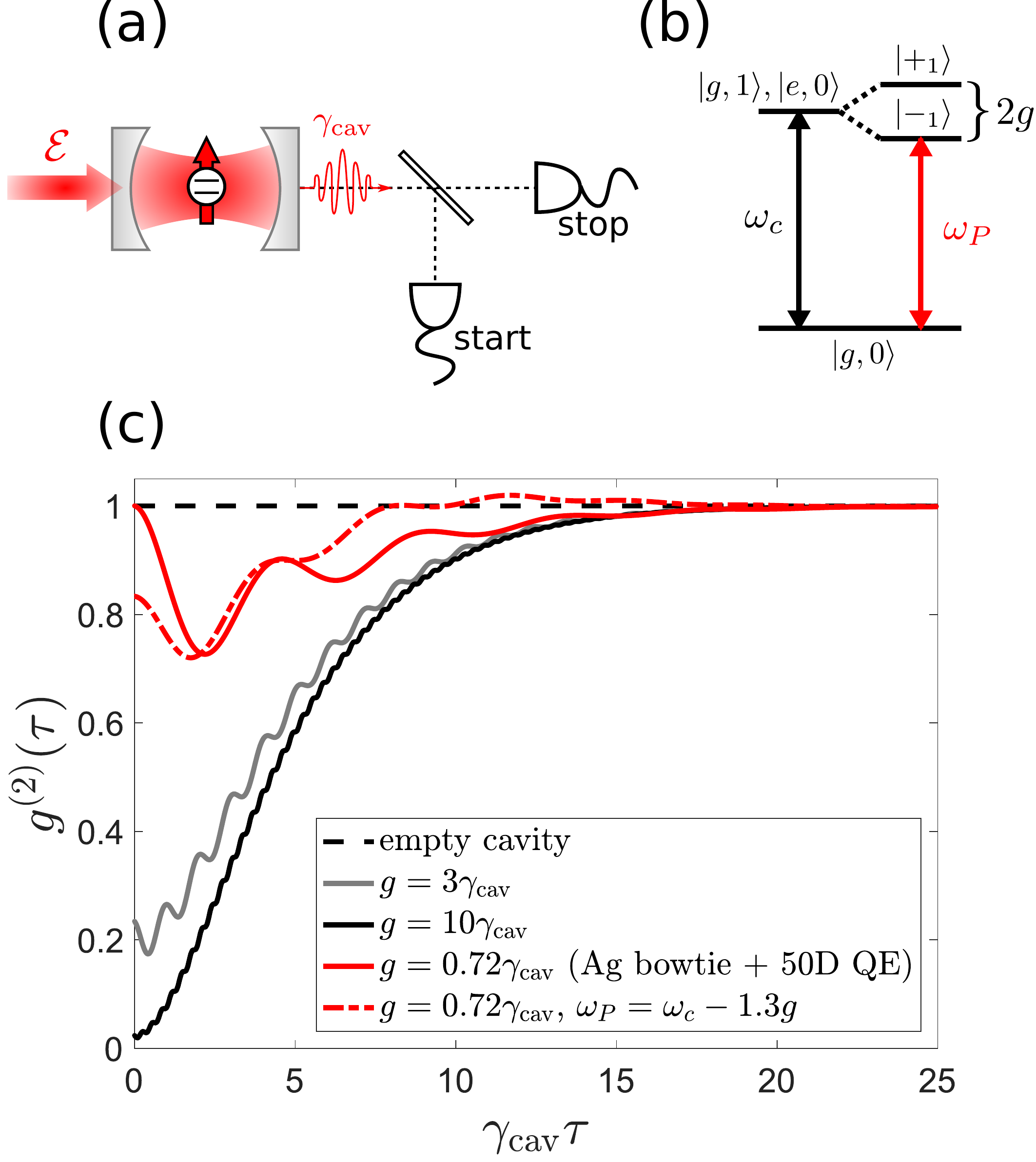}
\caption{(a) Hanbury-Brown-Twiss setup: a cavity containing a QE is coherently pumped with the laser Rabi frequency ${\cal E} = 0.1\gamma_{\rm cav}$ at frequency $\omega_P$, and the output field is split into two beams hitting avalanche photodiodes. (b) Linkage pattern of the single photon Jaynes-Cummings ladder, with the ground state $|g,0\ket$ and the polariton states $|\pm_1\ket = \op{\Lambda}^\+_\pm|g,0\ket$. The lower polariton is pumped at frequency $\omega_P = \omega_c - g$. (c) $g^{(2)}(\tau)$ function for different $g/\gamma_{\rm cav}$ ratios. The red lines corresponds to the Ag bow-tie with a 50D.}
\label{g2_func}
\end{figure}

Finally, we compute the photon statistics for the same configuration using the $g^{(2)}(\tau)$ function, corresponding to a Hanbury-Brown-Twiss experimental setup (see Fig. \ref{g2_func}(a)):
\begin{align}
\label{g2}
g^{(2)}(\tau) = \frac{\bra\,\op{a}^\+(t)\op{a}^\+(t+\tau)\op{a}(t+\tau)\op{a}(t)\ket}{\bra\op{a}^\+(t)\op{a}(t)\ket\bra\op{a}^\+(t+\tau)\op{a}(t+\tau)\ket}.
\end{align}
This function is defined as the probability of detecting a photon at time $t+\tau$ while another photon was detected at time $t$, normalized by the probabilities of detecting a single photon at the same times. Non-classical features of light are revealed when $g^{(2)}(\tau) < 1$, and in particular, antibunching is observed when $g^{(2)}(0) < 1$. In that case, photons tend to be emitted one by one. Calculations of the $g^{(2)}(\tau)$ function are shown in Fig. \ref{g2_func}(c), for low pump (${\cal E} = 0.1\gamma_{\rm cav}$). The red lines show the calculation for the Ag bow-tie coupled to a 50D QE, and oscillations due to the QE-plasmon coupling $g$ arise in the statistics, suppressing the antibunching when the system is pumped at $\omega_P = \omega_c - g$. This is due to the small splitting observed between the upper and lower polariton states, and for larger splitting, it is seen that the oscillations are faster and averaged into an overall smooth behavior. However, if the pump is slightly red-shifted to the lower polariton $|-_1\ket$ (red dash-dotted line in Fig. \ref{g2_func}(c)), it is seen that $g^{(2)}(0) <1$ and therefore a small antibunching is observed. We emphasize that the calculation of the photon statistics is done assuming that it is possible to correlate the output signals. Nevertheless, the time scale observed here is of the order of $\gamma_{\rm cav}\tau \sim 5$, which is about 30 fs. Even though it is challenging to access $g^{(2)}(\tau)$ at this fast time scale, it is however interesting to underline the non-classical nature of the light emitted by a QE coupled to a nanoantenna. Another remark can be made about the photon statistics: in \eqref{g2}, we assumed that the output field from which we determine the photon statistics is composed only of contributions from the cavity mode. Using the input-output theory, we can write the output mode as:
\begin{align}
\op{a}_{\rm out} = \op{a}_{\rm in} + \sqrt{\gamma_{\rm cav}}\,\op{a} + \sqrt{\gamma_0}\,\hsig_-.
\end{align}
In general, $g^{(2)}(\tau)$ should be computed using $\op{a}_{\rm out}$. We neglected the contribution from the QE as we have $\gamma_{\rm cav}\gg\gamma_0$ in our situation, but in the Purcell regime it should be included. If this is done properly, the antibunching can be observed on the nanoseconds to picoseconds time scales.

\section{Conclusion}
To conclude, we have systematically analyzed various plasmonic nanoantennas in the context of the Rabi splitting with single quantum emitters. We established the general condition that can be used to estimate the feasibility of the strong coupling regime, and applied it to study the required dipole moment of a single emitter to realize the strong coupling regime. We have found that Ag and Au bow tie nanoantennas present the most favorable structures for attaining strong coupling with single emitters. The conditions seem to be especially viable for the case of single excitons of two-dimensional semiconductors such as WS{$_2$}. Finally, we demonstrated numerically the emergence of single photon nonlinearities with the saturation of a strongly coupled system made of a QE placed in the gap of a silver bow-tie nanoantenna, using a master equation formalism. Future research should be aimed at exploring even more beneficial nanoantenna geometries and addressing single excitons in two-dimensional materials. This opens the perspective of realizing broadband photon-photon interactions at the nanoscale.

\begin{acknowledgments}
The authors acknowledge financial support from Knut and Alice Wallenberg Foundation. T.S. and G.J. acknowledge financial support from Swedish Research Council (VR grant number: 2016-06059).
\end{acknowledgments}

\bibliography{SC_cond_sing_em_BR_v2.bbl} 
\end{document}